\documentclass[aps,prl,twocolumn,groupedaddress]{revtex4}
\usepackage{amsmath}
\usepackage{amsfonts}

\def\bF{\mathbf F}
\def\bE{\mathbf E}
\def\bH{\mathbf H}
\def\bJ{\mathbf J}
\def\bB{\mathbf B}
\def\bD{\mathbf D}

\def\bX{\mathbf X}
\def\bY{\mathbf Y}

\def\bx{\mathbf x}

\def\bK{\mathbf K}

\def\b0{\mathbf 0}

\begin{document}

\title{Essential state of the electromagnetic field and the double-slit experiment}
\author{Neil V. Budko}
\affiliation{Laboratory of Electromagnetic Research, Faculty of Electrical Engineering, Mathematics and Computer Science,
Delft University of Technology,
Mekelweg 4, 2628 CD Delft, The Netherlands}
\email{n.budko@ewi.tudelft.nl}
\author{Alexander B. Samokhin}
\affiliation{Department of Applied Mathematics, Moscow Institute of
Radio Engineering, Electronics, and Automatics (MIREA), Verndasky~av.~78, 117454, Moscow, Russian Federation}
\thanks{This research is supported by NWO (The Netherlands)}

\date{\today}

\begin{abstract}
A new class of generalized solutions related to the essential spectrum of linear Maxwell's equations is presented. The essential modes
are given in terms of normalized singular Weyl's sequences, whose square represents Dirac's delta functions in spatial and angular 
frequency domains. The action integral associated with essential modes is well-defined. We claim that these modes represent the collapsed state
of the electromagnetic field and, with some additional assumptions on the conservation of action, are suitable for 
describing the double-slit experiment in accordance with the orthodox point of view.
\end{abstract}

\pacs{41.20.-q, 42.25.-p}

\maketitle
Wave motion plays a fundamental role in both classical and quantum physics.
Mathematically, the propagation of classical (physical) waves as well as evolution of quantum-mechanical 
wavefunctions (probability amplitudes) are described by some kind of a linear 
wave equation. In quantum physics, however, wave motion is not everything. During the measurement 
wavefunction undergoes what is known as {\it wavefunction collapse}. This collapse is one of the phenomenological 
{\it postulates} of the orthodox quantum theory. It remains a postulate, since such collapses are believed to be 
mathematically incompatible with the linear wave motion in general. Probably, the most well-known and at the same time 
the most paradoxical case is the double-slit experiment with photons, where a spatially extended wave collapses 
into what seems to be a truly random point at the detector screen. A large collection of such points, however, accumulates into 
a classical interference pattern. 

Although, the standard quantum optics operates within the postulates
of the orthodox theory and provides excellent predictions
about everything that has to do with photons, many find the pragmatic orthodox viewpoint difficult 
to accept and look for some kind of a {\it process}, which would resemble the collapse. 
There are two possibilities in this respect. One is to find the explanation for collapse
within the quantum physics itself. For example, the decoherence approach \cite{Zurek2003} 
considers the interference of partially coherent wavefunctions entangled with the
environment. Although this provides a better understanding of the quantum-classical transition, 
supporters of the decoherence approach tend to dismiss the
orthodox collapse as an awkward and unnecessary theoretical construct.
The other possibility is to find the classical solution of a wave equation which resembles
the collapse, as done in the ``dynamical reduction'' program \cite{BassiGhirardi2003}. This program 
is largely based on the ideas from classical nonlinear wave motion. 
However, nonlinear focusing is never perfect, so that the idealistic orthodox collapse is 
dismissed within this approach as well. Moreover, nonlinear wave propagation
is a causal and deterministic process. Therefore, in addition to an artificial nonlinearity 
in the modified Schr{\"o}dinger equation the dynamical reduction program introduces a stochastic term as well.

Here, somewhat contrary to the growing consensus, we would like to defend the orthodox point of view,
namely, the idea of collapse.  
Ontological (interpretational) problems aside, the main shortcoming of this idea is
its postulated nature. Of course, we do not pretend here to know the general solution of this problem on an abstract level. 
However, in the case of the electromagnetic wave motion, we do seem to have found a unified description 
of the propagating and collapsed electromagnetic field, so that both causal electromagnetic waves and the 
random points at the detector come out as solutions of linear Maxwell's equations. 
This is the subject of our Letter. It is difficult to decide whether our approach is classical or quantum-mechanical.
Probably, it is neither. And, may be, it is only natural, since the problem lies somewhere 
at the vague border between quantum and classical. 

To accommodate for collapse one needs a perfectly localized natural
mode of the electromagnetic field, which solves linear Maxwell's equations
in a certain, possibly generalized, sense. Elsewhere we show that the required mode is related to the 
essential spatial spectrum of the electromagnetic scattering operator \cite{BudkoSamokhin2006a}, and
is given in terms of a singular Weyl's sequence \cite{BudkoSamokhin2006c}.  
The action integral and other quadratic physical quantities involving
the electromagnetic field are well-defined for this spatially localized mode. Thus, despite its 
generalized nature, we could not dismiss the obtained singular mode as a purely mathematical artifact
and looked for its physical interpretation.
It turns out that our work is closely related to experiments in photonics and plasmonics, which both 
exploit unusual resonant states of the electromagnetic field. Recently we have uncovered a 
link between the plasmonic resonance and the essential spectrum of
the electromagnetic scattering operator \cite{BudkoSamokhin2006b}.

However, a spatially localized mode, even a perfectly localized one, is not quite enough. Although, the connection is not immediately 
obvious, states of this kind are already embedded in the modern mathematical approach to quantum mechanics 
known under the collective name of Rigged Hilbert Space (RHS) \cite{Bohm1999}. 
Apparently, the generalized eigenfunctions 
of the essential spectrum mentioned above do not belong to the usual complete Hilbert space \cite{BudkoSamokhin2006c}. 
The RHS approach allows, in principle, the quantum
evolution to escape beyond the Hilbert space.
This idea, however, is still at odds with another unpopular part of the orthodox interpretation. 
Namely, suggestion that collapse happens {\it instantaneously} and is not the result of a 
causal evolution. Again, the general consensus is to replace the instantaneous orthodox quantum ``jumps'' with 
a possibly very fast, but continuous {\it process}, e.g. evolution of a quantum system into a decoherent state. 

Retaining the idea of a spatially localized resonant state,  
we suggest the following paradigm shift. First of all, we know that the discrete {\it spatial spectrum}
of the Maxwell operator corresponds to delocalized (global) eigenmodes of the electromagnetic field, either
confined within a resonator our rapidly decaying
at infinity. Whereas, as we show in 
\cite{BudkoSamokhin2006c} and here, the essential spatial spectrum corresponds to perfectly localized
and normalizable generalized modes. By analogy we introduce the discrete {\it temporal spectrum} 
(spectrum of the time-derivative operator)
corresponding to finite-duration 
processes, which have both the beginning and the end, e.g. a photon is first radiated and then absorbed.
The essential temporal spectrum 
corresponds to generalized quasi-harmonic ``processes'', properly normalized and perfectly localized in the angular frequency domain. 
Combining these observations we propose to distinguish between the ``normal'' and the ``essential''
{\it spatio-temporal} states of the field.

To be more precise mathematically, in the electromagnetic case we are talking about the ``Weyl'' rather than essential 
spectrum, as we apply the Weyl criterion \cite{HislopSigal1996} to the differential Maxwell's equations, and the operator in question 
is not selfadjoint. This however is not important for the following two reasons. First of all, the spatial 
essential (singular) modes we encounter here are, in fact, proven to be the generalized 
modes of the essential spectrum of the equivalent non-selfadjoint and non-normal integral operator 
\cite{BudkoSamokhin2006a}, \cite{BudkoSamokhin2006c}. Secondly, for our purposes the generalized (and normalized) nature 
of the modes captured by the Weyl criterion is much more important than the actual name of the spectrum 
they belong to. Consider the Maxwell equations
 \begin{align}
 \label{eq:Maxwell}
 \begin{split}
 -\nabla\times \bH +\partial_{t}\bD&=-\bJ,
 \\
 \nabla\times \bE +\partial_{t}\bB&=\b0,
 \end{split}
 \end{align}
with linear isotropic constitutive relations
$\bD(\bx,t)=\varepsilon(\bx,t)*\bE(\bx,t)$,
$\bB(\bx,t)=\mu(\bx,t)*\bH(\bx,t)$,
where $\varepsilon(\bx,t)$ and $\mu(\bx,t)$ are some continuous functions with finite 
spatial support, and star denotes the temporal convolution. In the matrix form these equations can be written as
 \begin{align}
 \label{eq:MaxwellMatrix}
 \left[
 \begin{array}{cc}
 \partial_{t}\varepsilon* & -\nabla\times \\ 
 \nabla\times & \partial_{t}\mu*
 \end{array}
 \right]
 \left[
 \begin{array}{c}
 \bE\\
 \bH
 \end{array}
 \right]
 =
 \left[
 \begin{array}{c}
 -\bJ\\
 \b0
 \end{array}
 \right].
 \end{align}
In operator notation we simply write ${\mathbb M}\bX = \bY$,
where ${\mathbb M}$ is the Maxwell operator, and $\bX$ and $\bY$ are six-vectors.
We introduce the Hilbert space with the following norm:
 \begin{align}
 \label{eq:Norm}
 \left\Vert \bX \right\Vert^{2} = \int_{-\infty}^{\infty}\int_{\bx\in{\mathbb R}^{3}}
 \left\vert\bX\right\vert^{2}
 \,{\rm d}\bx\,{\rm d}t.
 \end{align}
According to the Weyl's definition of the spectrum of a selfadjoint operator
${\mathbb M}$ one needs to find the sequence of functions $\{\bF_{n}\}$ with the following properties:
 \begin{align}
 \label{eq:Minimization}
 \left\Vert\bF_{n}\right\Vert&=1,
\;\;\;\; \lim\limits_{n\rightarrow\infty}\left\Vert{\mathbb M}\bF_{n}-\lambda\bF_{n}\right\Vert=0
 \end{align}
If the Weyl sequence that satisfies these equations for some $\lambda$ has no convergent subsequence on the pertaining 
Hilbert space, then such $\lambda$ is in the essential spectrum. If for some other $\lambda$ the Weyl sequence does converge to 
a function from the Hilbert space, then this function is an eigenfunction, and $\lambda$ is an eigenvalue of ${\mathbb M}$.
With some reservations, which we mention earlier, we shall apply this definition to the Maxwell operator. 
The singular modes of the volume integral operator found in \cite{BudkoSamokhin2006c} are the following 
vector-valued functions:
\begin{align}
\label{eq:DefPsi}
\begin{split}
&\Psi(\alpha,\bx,\bx_{\rm c})=
\\
&\left(\frac{2}{3}\right)^{1/2}\pi^{-3/4}\alpha^{5/4}(\bx-\bx_{\rm c}) 
\exp\left(-\frac{\alpha}{2}\vert\bx-\bx_{\rm c}\vert^{2}\right),
\end{split}
\end{align}
where $\bx,\bx_{\rm c}\in{\mathbb R}^{3}$ and $\alpha\ge 0$ is the sequence parameter. 
The properties of these functions are summarized in Theorem~2.1 of \cite{BudkoSamokhin2006c}. 
The most important are the following two:
 \begin{align}
 \label{eq:NormalizedPsi}
 \int_{\bx\in{\mathbb R}^{3}}\left\vert\Psi(\alpha,\bx,\bx_{\rm c})\right\vert^{2}\,{\rm d}\bx&=1,
 \\
 \label{eq:Sifting}
 \lim\limits_{\alpha\rightarrow\infty}\int_{\bx\in{\mathbb R}^{3}}
 f(\bx)\left\vert\Psi(\alpha,\bx,\bx_{\rm c})\right\vert^{2}\,{\rm d}\bx &= f(\bx_{\rm c}),
 \end{align}
which is why we refer to singular modes as the square root of the delta function. 
As we intend to take into account the temporal spectrum, we shall consider 
a similar sequence of scalar functions:
 \begin{align}
 \label{eq:DefPhi}
 \begin{split}
 &\phi(\beta,t,\omega_{\rm c})=
 \\
 &i\sqrt{2}\pi^{-1/4}\beta^{-3/4}t
 \exp\left(-\frac{1}{2\beta}t^{2}-i\omega_{\rm c}t\right).
 \end{split}
 \end{align}
The Fourier transform of these functions is
 \begin{align}
 \label{eq:DefPhiOmega}
 \begin{split}
 &\hat{\phi}(\beta,\omega,\omega_{\rm c})=
 \\
 &\sqrt{2}\pi^{-1/4}\beta^{3/4}(\omega-\omega_{\rm c})
 \exp\left(-\frac{\beta}{2}\vert\omega-\omega_{\rm c}\vert^{2}\right),
 \end{split}
 \end{align}
and, as can be verified using the integration techniques developed in \cite{BudkoSamokhin2006c}, it is normalized and 
also represents the square root of the Dirac delta function, but now in $\omega$-domain, i.e.,
 \begin{align}
 \label{eq:NormalizedPhi}
 \int_{-\infty}^{\infty}\left\vert\hat{\phi}(\beta,\omega,\omega_{\rm c})\right\vert^{2}\,{\rm d}\omega&=1,
 \\
 \label{eq:SiftingPhi}
 \lim\limits_{\beta\rightarrow\infty}\int_{-\infty}^{\infty}
 g(\omega)\left\vert\hat{\phi}(\beta,\omega,\omega_{\rm c})\right\vert^{2}\,{\rm d}\omega &= g(\omega_{\rm c}).
 \end{align}
The product of functions (\ref{eq:DefPsi}) and (\ref{eq:DefPhi}) can be used to generate the essential 
modes of the complete Maxwell's operator, i.e.,
 \begin{align}
 \label{eq:EssentialModes}
 \bF^{\rm e}=
 \left[
 \begin{array}{c}
 \phi\Psi\\
 \b0
 \end{array}
 \right],
 \;\;\;\;\;\;
 \bF^{\rm m}=
 \left[
 \begin{array}{c}
 \b0\\
 \phi\Psi
 \end{array}
 \right]. 
 \end{align}
Indeed, substituting $\bF^{\rm e}$ and $\bF^{\rm m}$ in (\ref{eq:Minimization}) we eventually arrive at
 \begin{align}
 \label{eq:CheckFe}
 \lim\limits_{\alpha,\beta\rightarrow\infty}
 \left\Vert{\mathbb M}\bF^{\rm e}-\lambda^{\rm e}\bF^{\rm e}\right\Vert^{2}&=
 \left\vert\lambda^{\rm e}+i\omega_{\rm c}\hat{\varepsilon}(\bx_{\rm c},\omega_{\rm c})\right\vert^{2},
 \\
 \label{eq:CheckFm}
 \lim\limits_{\alpha,\beta\rightarrow\infty}
 \left\Vert{\mathbb M}\bF^{\rm m}-\lambda^{\rm m}\bF^{\rm m}\right\Vert^{2}&=
 \left\vert\lambda^{\rm m}+i\omega_{\rm c}\hat{\mu}(\bx_{\rm c},\omega_{\rm c})\right\vert^{2},
 \end{align}
meaning that $-i\omega_{\rm c}\hat{\varepsilon}(\bx_{\rm c},\omega_{\rm c})$ and 
$-i\omega_{\rm c}\hat{\mu}(\bx_{\rm c},\omega_{\rm c})$, where $\omega_{\rm c}\in{\mathbb R}$ and 
$\bx_{\rm c}\in{\mathbb R}^{3}$, are in the essential (or Weyl's) spectrum of the Maxwell operator.

Mathematical suitability of the essential state represented by one or more essential modes 
for describing the process of photon detection should be apparent by now.
Indeed, the photon is detected at a single point of the detector-screen, which is $\bx_{\rm c}$ in our case,
but also transfers a discrete portion of energy, which in our case may be associated with the angular frequency 
$\omega_{\rm c}$. This, however, is a tricky question, as we do not know, if any energy can be assigned
to the essential mode at all. 

What are the mathematical and physical conditions on the excitation of essential modes? 
From the Maxwell equations we deduce that $\bF^{\rm e}$ and $\bF^{\rm m}$
can only be excited by either induced or external currents with generalized components of the form $-\varepsilon_{0}\partial_{t}\bF^{\rm e}$
or $-\mu_{0}\partial_{t}\bF^{\rm m}$. This is applicable to both macroscopic and microscopic electrodynamics. 
In the macroscopic linear case with induced currents $\bJ^{\rm ind}=\partial_{t}[(\varepsilon*)-\varepsilon_{0}]\bE$
and $\bK^{\rm ind}=\partial_{t}[(\mu*)-\mu_{0}]\bH$
this leads to the condition on the essential resonance, i.e. 
$\lambda^{\rm e}=-i\omega_{\rm c}\hat{\varepsilon}(\bx_{\rm c},\omega_{\rm c})=0$
and $\lambda^{\rm m}=-i\omega_{\rm c}\hat{\mu}(\bx_{\rm c},\omega_{\rm c})=0$.

We notice another important similarity with the orthodox collapse, as essential states cannot causally {\it evolve} out of
an initially normal state, even in principle. If this could happen, then there would exist a sequence of
Hilbert space currents causally related to the sequence of fields, both converging to an essential state. 
However, this is not possible by the very definition of the essential state.
Hence, the normal~$\rightarrow$~essential state transformation is a non-causal ``jump-like'' event postulated by the
orthodox theory. 

Thus we cannot say much about the normal~$\rightarrow$~essential state transformation at the level of
the fields. If, however, the field is squared and integrated, then both the normal and the essential states
can be treated on equal footing. Note that a higher power (e.g. cube) of an essential mode gives an infinity, 
while the lower power produces a zero \cite{Colombeau1992}.
We shall approach the problem of the transformation in the following way.
Suppose that the normal state does not evolve into, but is simply {\it replaced} 
by an essential state. An equation describing such an event can take various forms and as we show 
below is equivalent to the conservation of the electromagnetic-field action in the presence of essential 
modes.

Assuming that the normal~$\rightarrow$~essential replacement transformation takes place, the practical 
problems one should try to solve in relation to the double-slit experiment are: 
the spatial locations of the essential modes, their angular frequencies, and the precise amount 
of action they carry (action is the most elementary squared-and-integrated quantity involving fields).
But first we note an important degree of freedom, a kind of non-uniqueness, associated with the essential modes. Namely,
condition for the essential resonance $\lambda^{\rm e}=-i\omega_{\rm c}\hat{\varepsilon}(\bx_{\rm c},\omega_{\rm c})=0$,
if satisfied at all,
will, in general, be satisfied not just at a single point in space, but along a hyper-surface or even
a hyper-volume. If a zero permittivity sounds like something suspiciously unphysical, one is advised to
think of essential induced currents instead, without any reference to the macroscopic constitutive
relations.
In any case, we shall assume that the measurement apparatus is prepared in such a way that the 
conditions on the excitation of essential modes are satisfied at the entire 
detector screen and for some
range of angular frequencies. From this it follows that certain expressions containing a 
{\it single} squared and integrated essential mode will
be invariant with respect to the location $\bx_{\rm c}$ and frequency $\omega_{\rm c}$ as long as 
this location is somewhere at the detector screen and the frequency is in the prescribed range.

Indeed, let $\bE_{\rm nr}(\bx,t)$ and $\bE_{\rm es}(\bx,t)$ denote
the electric fields in the normal and essential states, respectively, undergoing the replacement 
transformation. For $\varepsilon_{0}\vert\bE_{\rm es}\vert^{2}=C \vert\bF^{\rm e}\vert^{2}$ consider 
the following basic conservation law: 
\begin{align}
 \label{eq:ConserveSingle}
 &\varepsilon_{0}\int_{-\infty}^{\infty}\int_{\bx\in{\mathbb R}^{3}}
 \left\vert\hat{\bE}_{\rm nr}(\bx,\omega)\right\vert^{2}
 \,{\rm d}\bx\,{\rm d}\omega =
 \\ \nonumber
 &C \lim\limits_{\alpha,\beta\rightarrow\infty}\int_{-\infty}^{\infty}\int_{\bx\in{\mathbb R}^{3}}
 \left\vert\hat{\phi}(\beta,\omega,\omega_{\rm c})\Psi(\alpha,\bx,\bx_{\rm c})\right\vert^{2}
 \,{\rm d}\bx\,{\rm d}\omega
 =C,
 \end{align}
where $C$ is a fixed dimensional constant, which can and, probably, should be chosen as $\hbar$
--  the Plank constant, since it represents the elementary portion of action. 
To properly justify the above law one should, perhaps, look into the electromagnetic field part 
of the lagrangian $\mu_{0}\vert\bH\vert^{2}-\varepsilon_{0}\vert\bE\vert^{2}$ 
(which is the same in classical and quantum electrodynamics). For instance, we see that 
the simultaneous addition of $\varepsilon_{0}\vert\bE_{\rm es}\vert^{2}=C \vert\bF^{\rm e}\vert^{2}$
and $\mu_{0}\vert\bH_{\rm es}\vert^{2}=C \vert\bF^{\rm m}\vert^{2}$  
does not change the value of the action integral.
Another possibility is to consider not only the smooth variations of the action integral as
we do when we recover the classical equations of motion, including the Maxwell equations, 
but also its ``essential'' variations. This, however, is beyond the scope of the present Letter.

Obviously, the basic conservation law (\ref{eq:ConserveSingle}) does not depend 
on $\bx_{\rm c}$ and $\omega_{\rm c}$.
However, if we are to describe the double-slit experiment, the location and frequency 
of a number of modes has to be distributed in accordance with the intensity (squared amplitude) 
of the field in the normal state -- the Born statistical postulate.
We can show that the Born postulate is equivalent to a more general conservation law
associated with the normal-essential state transformation.
Let each essential mode $\bF^{\rm e}$ carry a finite portion of
action as in equation (\ref{eq:ConserveSingle}). Hence, position $\bx_{\rm c}$ and 
frequency $\omega_{\rm c}$ of each particular essential mode 
are, in general, not defined (arbitrary). We could say that they are ``random'', but this assumption does not
seem to be necessary.
Further, for {\it any} function $u(\bx,\omega)$, let the essential states satisfy 
the following generalized conservation law:
 \begin{align}
 \label{eq:ConserveGeneral}
 \begin{split}
 &\varepsilon_{0}\int_{-\infty}^{\infty}\int_{\bx\in{\mathbb R}^{3}}
 u(\bx,\omega)\left\vert\hat{\bE}_{\rm nr}(\bx,\omega)\right\vert^{2}
 \,{\rm d}\bx\,{\rm d}\omega =
 \\
 &\varepsilon_{0}\int_{-\infty}^{\infty}\int_{\bx\in{\mathbb R}^{3}}
 u(\bx,\omega)\left\vert\hat{\bE}_{\rm es}(\bx,\omega)\right\vert^{2}
 \,{\rm d}\bx\,{\rm d}\omega,
 \end{split}
 \end{align}
where expression on the right contains all
the necessary limits over Weyl's sequences. A similar law must hold for the magnetic fields.
Our goal is to determine the distributions of $\bx_{\rm c}$ and $\omega_{\rm c}$ for the resulting 
essential state.
Suppose that the normal state is transformed into $N$ essential modes, then
 \begin{align}
 \nonumber
 &\varepsilon_{0}\int_{-\infty}^{\infty}\int_{\bx\in{\mathbb R}^{3}}
 u(\bx,\omega)\left\vert\hat{\bE}_{\rm nr}(\bx,\omega)\right\vert^{2}
 \,{\rm d}\bx\,{\rm d}\omega =
 \\ \nonumber
 &\lim\limits_{\alpha,\beta\rightarrow\infty}NC\int_{-\infty}^{\infty}\int_{\bx\in{\mathbb R}^{3}}
 u(\bx,\omega)
 \int_{-\infty}^{\infty}\int_{\bx_{\rm c}\in{\mathbb R}^{3}}
 f(\bx_c)g(\omega_{\rm c})\times
 \\ \nonumber
 &\left\vert\hat{\phi}(\beta,\omega,\omega_{\rm c})\Psi(\alpha,\bx,\bx_{\rm c})\right\vert^{2}
 \,{\rm d}\bx_{\rm c}\,{\rm d}\omega_{\rm c}\,{\rm d}\bx\,{\rm d}\omega=
 \\  \label{eq:ConserveGeneralModes}
 &NC\int_{-\infty}^{\infty}\int_{\bx\in{\mathbb R}^{3}}
 u(\bx,\omega)
 f(\bx)g(\omega)\,{\rm d}\bx\,{\rm d}\omega,
 \end{align}
where  $f(\bx_c){\rm d}\bx_{c}$ and $g(\omega_{\rm c}){\rm d}\omega_{c}$ are the fractions
of essential states with parameters within the corresponding subvolume and interval, i.e., 
$f(\bx_{\rm c})$ and $g(\omega_{\rm c})$ are the distributions to be found. 
For arbitrary functions $u(\bx,\omega)$ this equality holds if and only if
\begin{align}
 \label{eq:BornRule}
 f(\bx)g(\omega)= \frac{\varepsilon_{0}}{NC}\left\vert\hat{\bE}_{\rm nr}(\bx,\omega)\right\vert^{2}.
 \end{align}
In other words, although for each particular essential mode parameters $\bx_{\rm c}$ 
and $\omega_{\rm c}$ are not defined, 
their {\it distribution} is equal to the normalized squared amplitude of the electromagnetic field 
in the normal state, i.e. for large $N$ the essential modes should form an interference pattern.

The electromagnetic field can, apparently, exist in two distinct states.
One, related to the discrete space-time spectrum, is the causal state, i.e., the classical 
electromagnetic wave.
The other state, related to the essential space-time spectrum, represents the collapse
of the electromagnetic field into generalized electric/magnetic modes with undefined position
and angular frequency. These modes are given in terms of singular Weyl sequences that
do not converge (in norm) to any function on the Hilbert space.
The two states cannot causally evolve into each other.
However, they may, under right circumstances (e.g. essential resonance), undergo a replacement 
transformation obeying certain conservation laws. In particular, we have made two plausible assumptions about the 
normal~$\rightarrow$~essential state transformation. First, each essential mode can carry (absorb) only a finite 
portion of action, which, in fact, is an extension of the Plank law.
Second, these transformations must obey a generalized conservation law. Under these assumptions the
Born statistical postulate is uniquely recovered. Thus, it would be interesting to find a proper justification
for the mentioned generalized conservation law. If we are right about the nature of collapse, then there should 
be a higher-level equation which governs the {\it rate} of collapse. Such an equation would, probably, involve action integrals
and describe the overall ratio of normal and essential states and the change of this ratio with time.

\end{document}